\documentclass[aps,prl,twocolumn,superscriptaddress,floatfix]{revtex4-2}

\usepackage[english]{babel}
\usepackage[dvips]{graphicx}
\usepackage{graphics}
\usepackage{amsfonts}
\usepackage{amsmath}
\usepackage{amssymb}
\usepackage{exscale}
\usepackage{eufrak}
\usepackage{afterpage}
\usepackage{upgreek}
\usepackage{color}	
\usepackage{braket}	
\usepackage{multirow}	
\usepackage{enumitem}
\usepackage{balance}

\usepackage{epstopdf}

\usepackage{miller}	

\begin{document}

\preprint{Experimental Physics}

\title{Experimental observation and spin texture of Dirac node arcs in tetradymite topological metals}

\author{J.~Dai}
\thanks{ji.dai@epfl.ch}
\altaffiliation[Present affiliation: ]{Institute of Physics and Lausanne Centre for Ultrafast Science (LACUS), 
			\'Ecole Polytechnique F\'ed\'erale de Lausanne, CH-1015 Lausanne, Switzerland}
\affiliation{Universit\'e Paris-Saclay, CNRS,  Institut des Sciences Mol\'eculaires d'Orsay, 
			91405, Orsay, France}
			
\author{E.~Frantzeskakis}
\thanks{emmanouil.frantzeskakis@u-psud.fr}
\affiliation{Universit\'e Paris-Saclay, CNRS,  Institut des Sciences Mol\'eculaires d'Orsay, 
			91405, Orsay, France}

\author{N.~Aryal}
\altaffiliation[Present affiliation: ]{Condensed Matter Physics and Materials Science Division, 
			Brookhaven National Laboratory, Upton, New York 11973, USA}
\affiliation{National High Magnetic Field Laboratory, Florida State University, United States of America}
\affiliation{Department of Physics, Florida State University, United States of America}

\author{K.-W.~Chen}
\altaffiliation[Present affiliation: ]{Department of Physics, University of Michigan, 
			Ann Arbor, MI, USA}
\affiliation{National High Magnetic Field Laboratory, Florida State University, United States of America}
\affiliation{Department of Physics, Florida State University, United States of America}

\author{F.~Fortuna}
\affiliation{Universit\'e Paris-Saclay, CNRS,  Institut des Sciences Mol\'eculaires d'Orsay, 
			91405, Orsay, France}
			

\author{J.~E.~Rault}
\affiliation{Synchrotron SOLEIL, L'Orme des Merisiers, Saint-Aubin-BP48, 91192 Gif-sur-Yvette, France}

\author{P.~Le~F\`evre}
\affiliation{Synchrotron SOLEIL, L'Orme des Merisiers, Saint-Aubin-BP48, 91192 Gif-sur-Yvette, France}

\author{L.~Balicas}
\affiliation{National High Magnetic Field Laboratory, Florida State University, United States of America}
\affiliation{Department of Physics, Florida State University, United States of America}

\author{K.~Miyamoto}
\affiliation{Hiroshima Synchrotron Radiation Center (HSRC), Hiroshima University, 2-313 Kagamiyama,
Higashi-Hiroshima 739-0046, Japan}

\author{T.~Okuda}
\affiliation{Hiroshima Synchrotron Radiation Center (HSRC), Hiroshima University, 2-313 Kagamiyama,
Higashi-Hiroshima 739-0046, Japan}

\author{E.~Manousakis}
\affiliation{National High Magnetic Field Laboratory, Florida State University, United States of America}
\affiliation{Department of Physics, Florida State University, United States of America}
\affiliation{Department of Physics, National and Kapodistrian University of Athens, 
Panepistimioupolis, Zografos, 157 84 Athens, Greece}

\author{R.~E.~Baumbach}
\affiliation{National High Magnetic Field Laboratory, Florida State University, United States of America}
\affiliation{Department of Physics, Florida State University, United States of America}

\author{A.~F.Santander-Syro}
\thanks{andres.santander-syro@u-psud.fr}
\affiliation{Universit\'e Paris-Saclay, CNRS,  Institut des Sciences Mol\'eculaires d'Orsay, 
			91405, Orsay, France}
\date{\today}

\begin{abstract}
	We report the observation of a non-trivial spin texture in Dirac node arcs, 
	novel topological objects formed when Dirac cones of massless particles 
	extend along an open one-dimensional line in momentum space. 
	We find that such states are present in all the compounds of the tetradymite 
	M$_2$Te$_2$X family (M$=$Ti, Zr or Hf and X$=$P or As), 
	regardless of the weak or strong character of the topological invariant. 
	The Dirac node arcs in tetradymites are thus the simplest possible, textbook example, 
	of a type-I Dirac system with a single spin-polarized node arc.
\end{abstract}
\maketitle

During the last decade various topological phases of matter
such as the quantum spin Hall effect~\cite{Kane2005, Konig2007}, 
topological insulators~\cite{Hasan2010,Fu2007a}, topological superconductors~\cite{Qi2011}, 
topological crystalline insulators~\cite{Fu2011}, 
or Dirac and Weyl semimetals~\cite{Soluyanov2015, Liu2014, Xu2015a, Xu2015b},
have provoked an immense interest of the scientific community. 
Scientific excitement in topological matter stems 
from its high potential in novel applications such as building blocks of quantum information 
in the form of Majorana zero modes~\cite{Stern2013, Lahtinen2017}, 
spin generators in spintronic circuits~\cite{Pesin2012, Khang2018} 
and optoelectronic nanodevices~\cite{Yue2016}.

Dirac cones, the simplest form of topological states, 
can have variable dimensionality in momentum space ranging from 1D to 3D, 
and display different topological characteristics giving rise to weak and strong topological states. 
Moreover, when the Dirac cone extends along an open 1D line in momentum space,
the series of adjacent Dirac points form so-called ``Dirac node arcs''~\cite{Wu2016,Takane2016}. 
Each variation in dimensionality, topological characteristics and $k$-space fingerprint 
of the Dirac point (i.e. a single node vs. a 1D line or arc) 
defines a new class of topological matter: quantum spin Hall insulators 
for 1D states~\cite{Konig2007, Knez2011}, 
strong or weak topological insulators for 2D states~\cite{Xia2009, Chen2009, Noguchi2019}, 
Dirac and Weyl semimetals for 3D states~\cite{Liu2014, Xu2015a, Neupane2014, Xu2015b}, 
nodal-line semimetals for Dirac nodal lines~\cite{Burkov2011,Bian2016,Schoop2016}.

There is however a new family of topological metals that is predicted to combine 
many of the above special features. This is the tetradymite family M$_{2}$Te$_{2}$X 
(with M$=$Ti, Zr or Hf and X$=$P or As), in which both strong and weak topological surface states 
have been theoretically predicted~\cite{Chen2016, Ji2016, Chen2018, Hosen2018}, 
interestingly accompanied by Dirac node arcs in one of its members~\cite{Hosen2018}. 
In this work, we employ angle- and spin-resolved photoemission spectroscopy (ARPES and SARPES) 
to experimentally demonstrate the crucial -yet missing- element 
that such Dirac node arcs are \emph{spin-polarized}, showing a non-trivial spin-texture,
and hence they meet all the essential requirement to be classified as 
\emph{topological Dirac node arcs}. 
Moreover, our results prove that \emph{all} compounds 
of the family share the exotic feature of a Dirac node arc. 
Compared to Dirac node arcs observed in 
topological line-node semimetals~\cite{Takane2016,Chen2017},
or to multiple node-arcs in type-I Dirac systems~\cite{Wu2016},
the unique Dirac node arc in the tetradymite M$_{2}$Te$_{2}$X compounds is thus
the simplest possible, textbook example, of a type-I Dirac system with a 
single spin-polarized node arc. 

\begin{figure*}[p!t!h]
   \centering
   \includegraphics[width=0.9\linewidth]{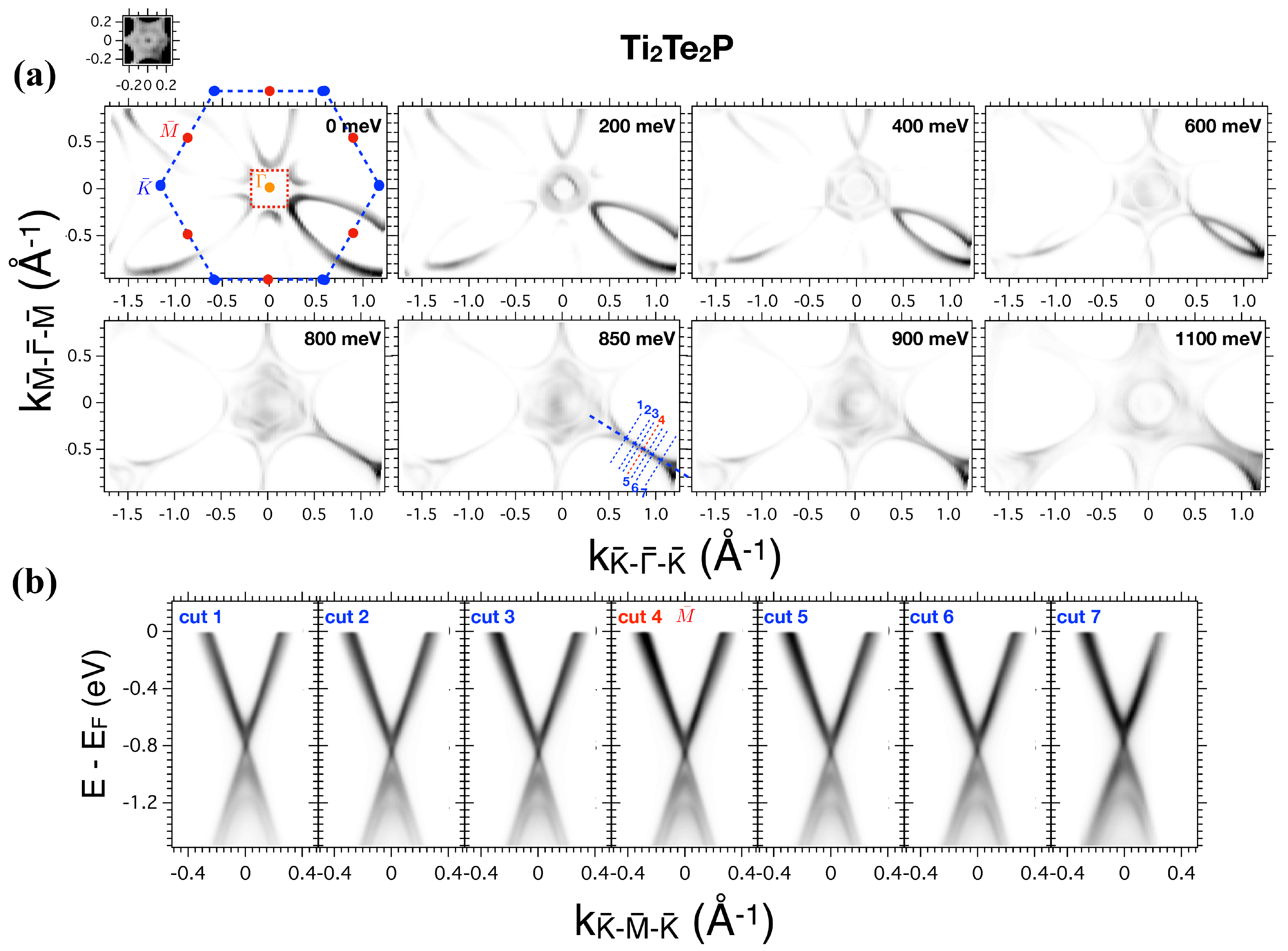}
   \caption{\footnotesize{(Color online)
   		   (a) Constant energy maps at various binding energies. 
   		   A linear feature appears at an approximate binding energy of 850~meV. 
   		   The dashed blue hexagon marks the borders of the surface-projected Brillouin zone. 
   		   The dashed red square indicates the area with saturated contrast shown in the inset
   		   above the top left panel.
   		   (b) Energy-momentum dispersion showing a persistent Dirac-like dispersion 
   		   along various $k$-paths marked by dashed lines in (a). 
   		   All data were collected 
   		   using photons of 50~eV and linear horizontal polarization. 
   		   The temperature was 6~K.
   		   }
   		}
   \label{normal_arpes}
\end{figure*} 

ARPES experiments were performed at the CASSIOPEE beamline of Synchrotron SOLEIL (France), 
and Spin-ARPES experiments were performed using the ESPRESSO machine 
at beamline 9B of the Hiroshima Synchrotron Radiation Center (HiSOR, Japan)~\cite{Okuda2011}.
Typical energy and angular resolutions were 15~meV and 0.25$^{\textmd{o}}$.  
The single crystals of M$_{2}$Te$_{2}$X (M$=$Ti, Zr or Hf and X$=$P or As)~\cite{Chen2016}
were cleaved \emph{in situ} at temperatures below 25~K
and pressure in the range of 10$^{-11}$~mbar, and were kept at those conditions
during the measurements.
The Supplementary Material provides complete technical details
about the sample growth, crystal structure, ARPES 
and Spin-ARPES measurements

Figure~1(a) presents the experimental constant energy contours of Ti$_{2}$Te$_{2}$P 
at various binding energies. The Fermi surface consists of six petal-like electron pockets 
centered at the $\overline{\textmd{M}}$ points of the surface Brillouin zone and 
a weaker hexagonal contour in the immediate vicinity of $\overline{\Gamma}$~\cite{Chen2018},
shown in the inset. 
The smallest point-like contour at $\overline{\Gamma}$ is residual intensity coming from a 
hole-like band whose maximum lies just below $E_{F}$ for Ti$_{2}$Te$_{2}$P \cite{Chen2018}. 
As the binding energy increases, the petal-like contours evolve into linear features
along the $\overline{\Gamma \textmd{M}}$ high symmetry lines. 
The energy-momentum dispersion along $\overline{\textmd{KMK}}$ reveals a Dirac cone 
that is typical for all compounds of the tetradymite family M$_{2}$Te$_{2}$X. 
Our previous first-principles calculations~\cite{Chen2018}
revealed the non-trivial origin of such Dirac-like state.
The binding energy of the Dirac point 
for Ti$_{2}$Te$_{2}$P is 0.85~eV, that is 200-300~meV lower than for other compounds 
of this family (see Fig.~\ref{EK_M}). Interestingly, as shown in panel (b) by the 
energy-momentum maps along $k$-paths parallel to $\overline{\textmd{KMK}}$, 
this Dirac-like dispersion is present all along the linear features of the constant energy map, 
with the Dirac point shifting to slightly lower binding energies 
as one moves away from $\overline{\textmd{M}}$ (see Fig.~\ref{ZTP_HTP_ZTA_arc}).
These results are in agreement with a previous study on Hf$_{2}$Te$_{2}$P~\cite{Hosen2018}, 
and they present a first indication that these linear features may correspond 
to topologically non-trivial Dirac node arcs.
The Supplementary Material presents additional data and analyses
for the linearly dispersing Dirac states in the vicinity of $\bar{M}$ along $\overline{\textmd{KMK}}$,
as well as for the Dirac node arcs, for Hf$_2$Te$_2$P, Zr$_2$Te$_2$P, Ti$_2$Te$_2$P and Zr$_2$Te$_2$As. 

Without any information on their spin texture, the linear features in the constant energy maps 
of the M$_{2}$Te$_{2}$X compounds (see Fig.~\ref{ZTP_HTP_ZTA_arc}), even if associated to a cone-like dispersion, 
cannot be unambiguously assigned to topologically non-trivial Dirac node arcs. 
Thus, a direct measurement of their spin polarization is necessary to elucidate if they correspond
to non-trivial states.
As we will see next, our SARPES data on Ti$_{2}$Te$_{2}$P and Hf$_{2}$Te$_{2}$P
reveal an appreciable spin polarization of the Dirac cones both along $\overline{\textmd{KMK}}$ 
and at parallel $k$-paths. As a result, they establish the linear features 
of the constant energy maps as topologically non-trivial 1D Dirac node arcs.
 
\begin{figure}[p!th] 
   \centering
   \includegraphics[clip, width=0.48\textwidth]{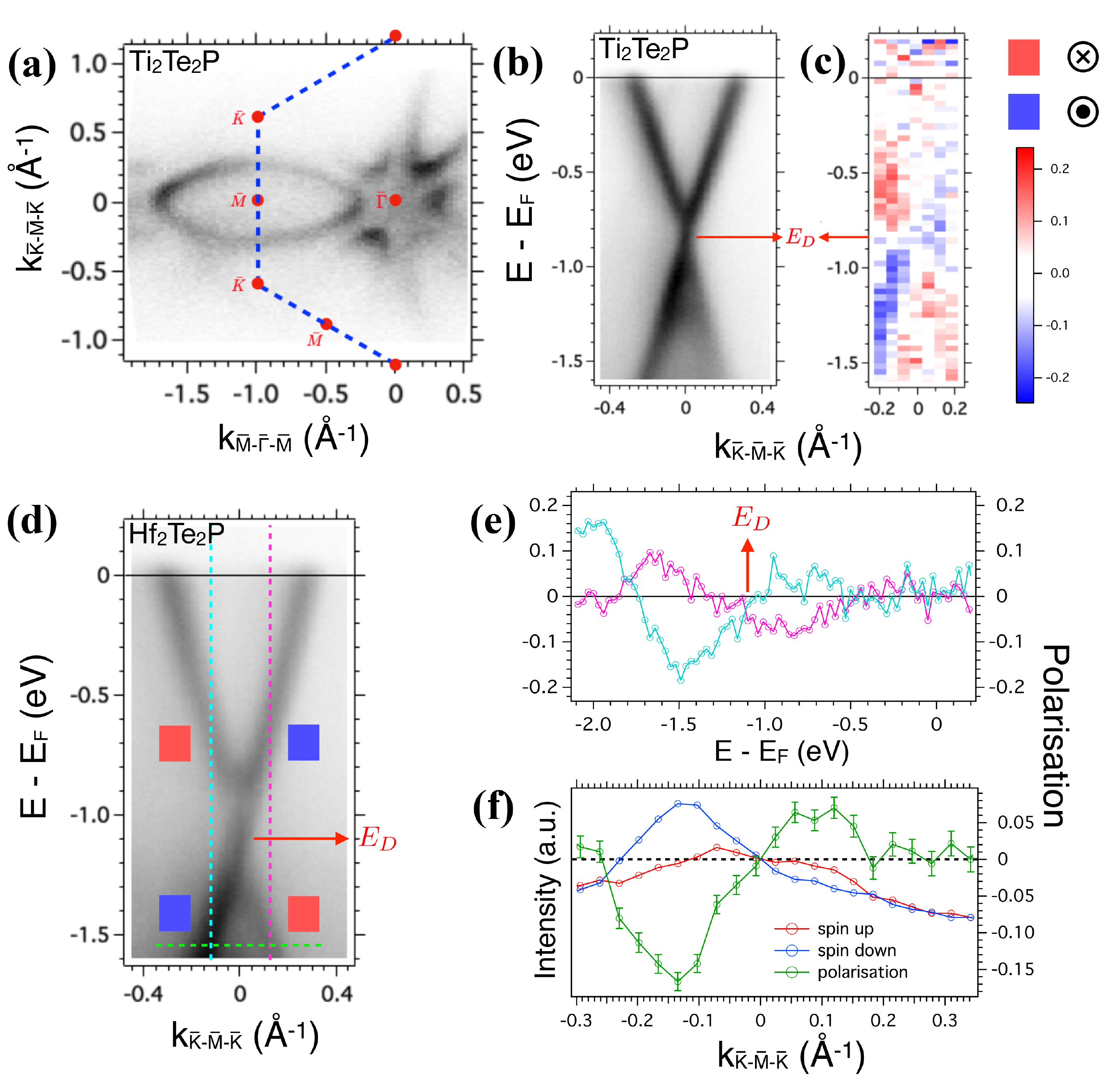} 
   \caption{\footnotesize{(Color online)
   		   (a) In-plane Fermi surface map of Ti$_{2}$Te$_{2}$P showing a sixfold symmetry. 
   		   (b,~c) Near-$E_{F}$ band structure of Ti$_{2}$Te$_{2}$P along the $\overline{\textmd{KMK}}$ 
   		   high-symmetry line with spin-integrated (b) and spin-resolved (c) ARPES, 
   		   showing a Dirac-like dispersion with clear spin polarization. 
   		   The color scale in (c) represents the sign and value 
   		   of the in-plane spin polarization 
   		   along $\overline{\Gamma \textmd{M}}$ with blue (red) pointing towards to 
   		   (away from) the reader.    		   
   		   (d) Near-$E_{F}$ band structure of Hf$_{2}$Te$_{2}$P 
   		   along the $\overline{\textmd{KMK}}$ high-symmetry line.
   		   Blue (red) squares indicate branches of the Dirac cone 
   		   with spin polarisation pointing towards to (away from) the reader.
   		   (e) Energy dependent spin polarization of Hf$_{2}$Te$_{2}$P 
   		   at the constant momenta indicated by the vertical light blue and magenta
   		   dashed lines in (c). 
   		   (f) Spin-resolved momentum distribution curves (blue, red) 
   		   and momentum dependent spin polarization (green) at the constant energy indicated 
   		   by the horizontal green dashed line in (d). Spin up/down means 
		   parallel/anti-parallel to $\overline{\Gamma \textmd{M}}.$
   		   The Dirac point at energy $E_D$ is shown by a red arrow in panels (b-e).
   		   All data were collected 
   		   with LH polarized photons of 55~eV. 
   		   The temperature was 25~K.
   		   }
   		}
   \label{M_point_spin}
\end{figure}

Figure~2 presents spin-integrated and spin-resolved ARPES results acquired 
using the ESPRESSO setup at HiSOR~\cite{Okuda2011} on Ti$_{2}$Te$_{2}$P, Figs.~2(a,~b), 
and Hf$_{2}$Te$_{2}$P, Figs.~2(c,~d,~e). 
The spin-integrated ARPES results on both compounds, panels (a,~b,~d), 
reproduce the main experimental features discussed before. 
Our goal is to establish the spin texture of the Dirac-like dispersion 
along $\overline{\textmd{KMK}}$ by measuring its in-plane spin component 
along the orthogonal $\overline{\Gamma \textmd{M}}$ direction. In the case of Ti$_{2}$Te$_{2}$P, 
the energy-dependent spin polarization has been acquired at different $k$-locations 
of the Dirac cone by measuring the spin-resolved energy distribution curves or EDCs 
(see the Supplementary Material). 
Panel (c) is a stack of the energy-dependent polarization curves 
where the color scale represents the sign of the in-plane spin polarization 
with blue (red) pointing towards to (away from) the reader. 
The Dirac-like dispersion is reproduced with a clear polarization reversal between the left and right sides 
of the cone. Moreover, our data reveal another reversal of the in-plane spin polarization 
between the top and bottom parts of the cone, 
in good agreement with results on Bi-based topological insulators~\cite{Hsieh2009}.

Similar conclusions can be drawn on the in-plane spin polarization of Hf$_{2}$Te$_{2}$P,
whose spin-integrated Dirac cone is shown in panel (d).
The energy dependent spin polarization at both sides of the Dirac cone,
shown in panel (e), 
has been acquired by means of spin-resolved EDCs. The polarization of each curve reverses 
above and below the Dirac point, while the two curves have opposite polarizations at a given binding energy. 
As expected, at the binding energy of the Dirac point ($1.1$~eV), 
the spin polarization is vanishingly small. Consistent information can be obtained with 
spin-resolved momentum distribution curves (MDCs), shown in panel (f). 
Here the change in the direction of the in-plane spin polarization component 
is tracked as a function of momentum at a fixed binding energy 
at the bottom part of the Dirac cone ($1.55$~eV). 
At this binding energy, the polarization changes from negative to positive as one passes 
from the left to the right branch of the Dirac cone. 
This is in perfect agreement with the data shown in panel (e) 
where the light blue (magenta) curve obtained at the left (right) side of the cone shows a 
negative (positive) polarization at a binding energy of $1.55$~eV. Based on the 
results presented in Fig.~2, we can experimentally confirm that the Dirac cone in compounds 
of the tetradymite family M$_{2}$Te$_{2}$X is spin polarized.

\begin{figure*}[pth]
   \centering
   \includegraphics[width=0.8\linewidth]{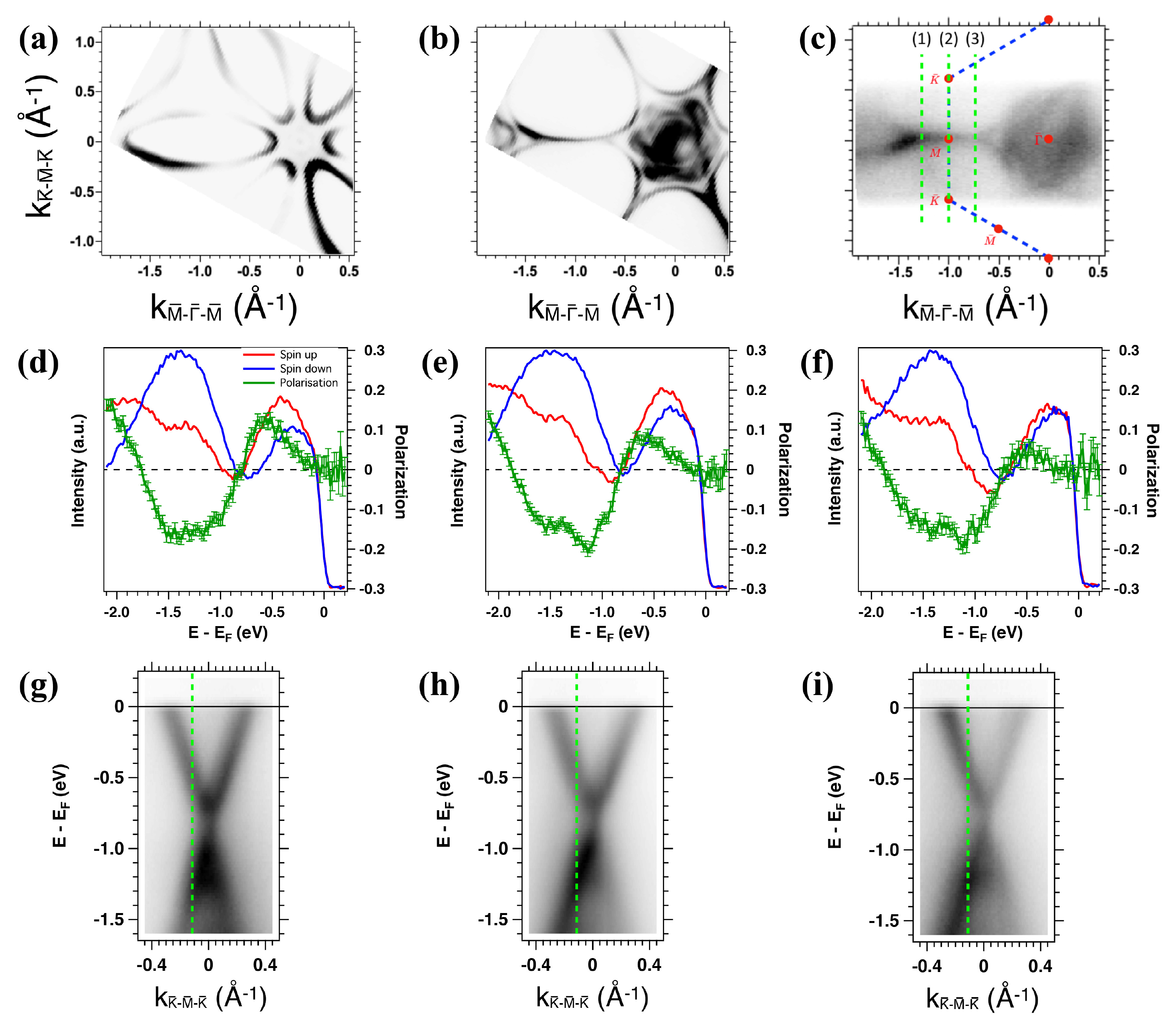} 
   \caption{\footnotesize{(Color online)
   		   (a), (b) Spin-integrated ARPES results obtained at synchrotron SOLEIL 
   		   with LH polarized photons of 50~eV at a temperature of 6~K showing the Fermi surface contours 
   		   of Ti$_{2}$Te$_{2}$P (a) and the existence of linear features at a binding energy 
   		   of 0.85~eV (b). 
   		   (c) Same as panel (b) but measured with LH photons of 55~eV at the ESPRESSO
   		   spin-resolved ARPES setup of beamline 9B at HiSOR. 
   		   Path (1) corresponds to data shown in panels (d) and (g), path (2) to panels (e) and (f), 
   		   and path (3) to panels (f) and (i). 
   		   (d)-(f) Spin-resolved EDCs (red, blue) and energy dependent in-plane spin polarization 
   		   parallel to $\overline{\Gamma \textmd{M}}$ (green) measured at the left branch of the Dirac cones 
   		   [panels (g)-(i)], themselves obtained at different $k$-locations along the Dirac node arc. 
   		   (g)-(i) Energy-momentum dispersion along the $k$-paths indicated in panel (c), with overlaid 
   		   vertical dashed lines showing the $k$-location of the spin-resolved EDCs shown 
   		   in panels (d)-(f). 
   		   Panels (c)-(f) were all measured with LH photons of 55~eV at a temperature of 25~K.
   		   }
   		}
   \label{Dirac_arc}
\end{figure*}

Having established the spin polarization of the Dirac cone at $\overline{\textmd{M}}$, 
we now turn our attention to the characteristics and the spin texture 
of the linear features observed in the constant energy maps of Fig. 1. 
Figs. 3(a) and 3(b) show once more how the petal contours of the Fermi surface of Ti$_{2}$Te$_{2}$P 
evolve into linear features at the binding energy of the Dirac point. 
Panel (c) presents another measurement of the constant energy contours 
at the Dirac point energy, 
acquired with the experimental setup at HiSOR 
right before spin-resolved measurements. 
Figures~3(d-f) present the spin-resolved EDCs and the resulting 
energy-dependent spin polarization acquired at the left side of the Dirac cones, panels (g-i),
themselves observed along three different $k$-paths
parallel to $\overline{\textmd{K}} \overline{\textmd{M}} \overline{\textmd{K}}$:
cuts (1) - (3) in panel (c). 
These data show that an in-plane spin-polarization exists not only along the 
$\overline{\textmd{K}} \overline{\textmd{M}} \overline{\textmd{K}}$
high-symmetry direction but also all along the Dirac node arc. 
We conclude that the Dirac cones along 
$\overline{\textmd{K}} \overline{\Gamma} \overline{\textmd{K}}$
not only share a common energy for their Dirac point, forming the node arc, 
but also that they exhibit an identical spin polarization. 
Our experimental results can therefore establish that the observed 
linear features at the Dirac point energy correspond indeed to topologically non-trivial Dirac node arcs. 

The aforementioned results confirm the presence of spin polarized electrons with a Dirac-like dispersion 
in one direction and a very large effective mass in the perpendicular direction 
(see also Fig.~\ref{ZTP_HTP_ZTA_arc}, Supplementary Material). 
Such qualitatively different behaviors may stem from the mixed contributions 
of $p$ ($\approx$ 60-70\%) and $d$ ($\approx$ 20-30\%) orbital states 
to the topological surface state at $\overline{\textmd{M}}$ 
(see Figs.~\ref{HTP_K_states} and~\ref{orbit_types}, Supplementary Material), 
as suggested for Ru$_2$Sn$_3$ \cite{gibson2014quasi}.
Provided that the Dirac cone could be tuned near $E_{F}$, compounds of this family could give rise 
to highly anisotropic 2D electron systems with spin polarized carriers. 
Our results prove that the existence of the Dirac node arcs is an inherent property 
of \emph{all} studied M$_{2}$Te$_{2}$X compounds, 
regardless of their topological character at $\overline{\textmd{M}}$, 
rather than being related to the weak topological character and the weak interlayer coupling 
of Hf$_{2}$Te$_{2}$P as suggested in Ref.~\onlinecite{Hosen2018}. 
After all, the Dirac states at $\overline{\textmd{M}}$ have a strong topological character 
for Zr$_{2}$Te$_{2}$P, Zr$_{2}$Te$_{2}$As and Ti$_{2}$Te$_{2}$P \cite{Chen2018, Ji2016}. 

The experimental observation of an in-plane spin polarization of the Dirac node arcs
in a direction normal to the arcs' crystal momentum
(i.e., parallel to $\overline{\Gamma \textmd{M}}$)
agrees with the main direction of the spin polarized vector in topological insulators 
and Rashba compounds~\cite{Hsieh2009,Sanchez2014}. 
The Supplementary Material shows additional data 
for the out-of-plane spin polarization, and discusses the magnitude of the
observed spin polarization.

Surprisingly, there is substantial spin polarization at energies $E-E_F < -1.5$~eV, 
i.e. below the lower branch of the Dirac cone (Figs.~2 and 3).
This observation has been reproduced 
in different experimental runs for all cleaved surfaces of both compounds studied here 
(Hf$_{2}$Te$_{2}$P and Ti$_{2}$Te$_{2}$P). On the other hand, it is not observed for compounds 
that do not belong to the M$_{2}$Te$_{2}$X that were studied with the same setup. 
Therefore, we believe that it is not due to an experimental artifact and it may indeed reveal 
the existence of spin polarized states at larger binding energies. This scenario is 
in agreement with the experimental observation (e.g. Fig.~1) and the theoretical 
prediction~\cite{Chen2018} of hole-like surface states at the same energy range. 
A possible explanation is the spin-polarized surface-confined states 
due to the Rashba-Bychkov effect, which have been repeatedly observed 
in the band structure of Bi$_{2}$Se$_{3}$, 
in the vicinity of both the upper and the lower branches 
of its Dirac cone~\cite{King2011,Bahramy2012}.

In conclusion, by means of spin-integrated and spin-resolved ARPES, 
we unambiguously proved the existence of type-I \emph{topological} Dirac node arcs 
in compounds of the M$_{2}$Te$_{2}$X family.
Our data showed bands with linear dispersion in one direction and a very large effective mass 
in the perpendicular direction. Our direct measurement of their spin polarization vector shows 
substantial in-plane spin polarization in the direction perpendicular to the crystal momentum
of the Dirac node arcs, 
all along the linear features in the constant energy contours. 
This helical arrangement of the electron spins is opposite for the upper and lower branches 
of the Dirac cone forming the arc.
Taken together, these observations establish the existence 
of topological Dirac node arcs in all studied compounds of the M$_{2}$Te$_{2}$X family 
regardless of their different topological characters.
An exciting perspective for future research would be to tune, by doping or pressure, 
the energy of the Dirac points in the M$_{2}$Te$_{2}$X family to the Fermi level.


\begin{acknowledgements}
  We thank F.~Bertran for assistance during ARPES measurements at CASSIOPEE (Synchrotron SOLEIL).
  ARPES work at ISMO was supported by public grants from the French National Research Agency (ANR), 
  project Fermi-NESt No ANR-16-CE92-0018.
  Experiments at HiSOR were performed under the approval of the 
  Program Advisory Committee (proposals 16BG014 and 17BU010).
  R.B. acknowledges support from the National Science Foundation through NSF/DMR1904361.
  L.B. is supported by DOE-BES through award DE-SC0002613.
\end{acknowledgements}

\section*{SUPPLEMENTARY MATERIAL}

\subsection*{Crystal structure}
\begin{figure}[p!h]
   \centering
   \includegraphics[clip, width=0.48\textwidth]{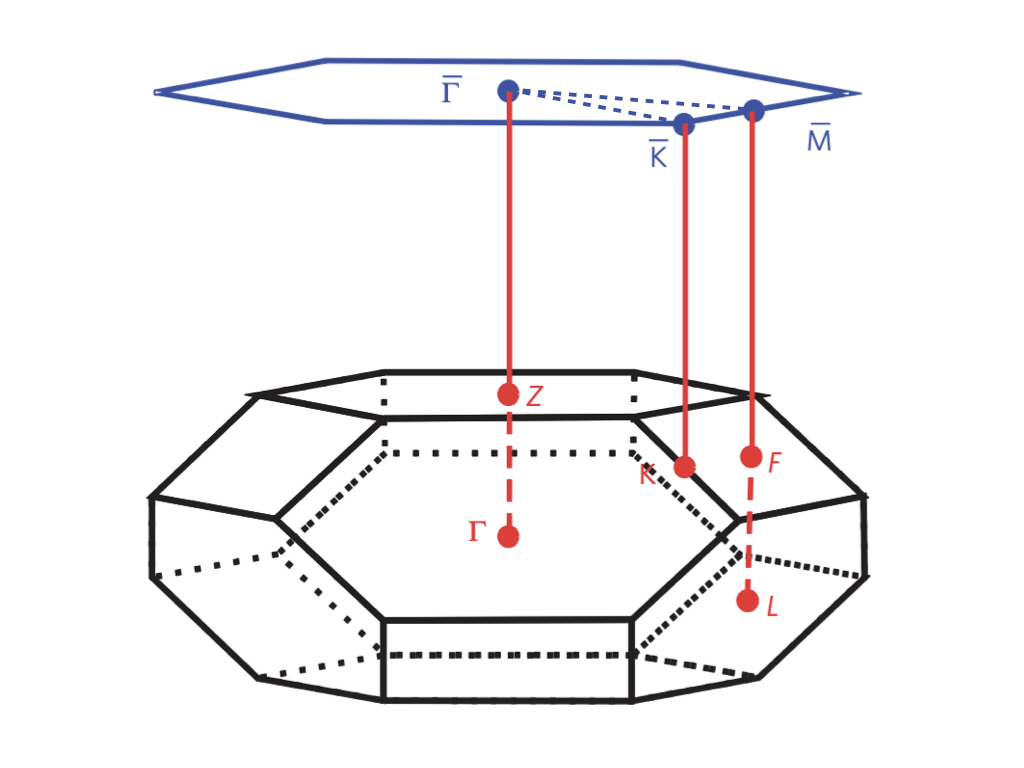} 
   \caption{The rhombohedral Brillouin zone projects onto a hexagonal surface Brillouin zone. 
   			The time-reversal invariant momentum (TRIM) points $\Gamma$, $Z$, $F$, $L$ 
   			are labeled as red dots.}
   \label{221_lattice_BZ}
\end{figure}
The crystal structure of M$_2$Te$_2$X (with M = Ti, Zr or Hf, and X = P or As) 
has the symmetry of the rhombohedral space group $R\bar{3}m$ (No.~166). 
The corresponding bulk Brillouin zone (rhombohedral) 
and the reduced surface Brillouin zone (hexagonal) are shown in Fig.~\ref{221_lattice_BZ}. 
The bulk TRIM points ($\Gamma$, $Z$, and 3 pairs of $F$ and $L$)
are labeled as red dots. Depending on the calculated parities, 
the topological surface states might be observed at $\bar{\Gamma}$ and/or $\bar{\textmd{M}}$ 
that correspond to the surface projection of the aforementioned bulk TRIM points.

\subsection*{Methods}
\textbf{Single crystal growth.} Single crystals of M$_{2}$Te$_{2}$X (M$=$Ti, Zr or Hf and X$=$P or As) 
were grown using the chemical vapor transport method that was described elsewhere \cite{Chen2016}. 
The precursor material consisted of stoichiometric polycrystals themselves prepared 
by the reaction of raw elements at 1000$^{\textmd{o}}$C for 24 hours. 
The resulting powders were sealed under vacuum with 3 mg/cm$^{3}$ of iodine in quartz tubes. 
The tubes were subsequently placed in a resistive tube furnace 
with a temperature gradient with 800$^{\textmd{o}}$C (source) and 900$^{\textmd{o}}$C (drain) for 21 days. 
Exfoliable single crystal specimens formed in the cold zone of the ampoule, 
having a hexagonal plate shape and the crystallographic $c$-axis oriented perpendicular 
to the plate face. Single crystals were cleaved in UHV conditions to expose an adsorbate-free surface 
right before the ARPES experiments.
\\

\textbf{Spin-integrated ARPES.} ARPES experiments were performed at the CASSIOPEE beamline 
of Synchrotron SOLEIL and at beamline 9B of the Hiroshima Synchrotron Radiation Center (HiSOR) 
using hemispherical electron analyzers with vertical and horizontal slits, respectively. 
Typical energy and angular resolutions were 15~meV and 0.25$^{\textmd{o}}$. 
The polarization of incoming photons was linear horizontal. 
Measurements were performed at 6~K (SOLEIL) or 25~K (HiSOR), 
under a typical pressure in the range of 10$^{-11}$~mbar.
\\

\textbf{Spin-resolved ARPES.} Spin-ARPES experiments were performed using the 
ESPRESSO machine at beamline 9B of the Hiroshima Synchrotron Radiation Center (HiSOR) \cite{Okuda2011}. 
Spin separation of the emitted photoelectrons is achieved by means of 
very low energy diffraction (VLEED) detectors. 
In the VLEED detector, photoelectrons are scattered by a Fe(001)-p(1$\times$1)-O 
film grown on a MgO(001) substrate \cite{Okuda2008}. The FeO film is magnetized 
by a pair of solenoids that impose two orthogonal magnetization axes in its surface. 
In order to measure the component of the spin normal to the Dirac node arc
(parallel to the $\overline{\Gamma \textmd{\textmd{M}}}$ line intersecting the arc),
the sample was aligned with the magnetization axis of the FeO film. 
A second VLEED detector oriented at 90$^{\textmd{o}}$ with respect to the first one allows 
for measurements of all three spin polarization components. 
The spin asymmetry ($A$) of the electron beam is measured by switching the magnetization of the of FeO film 
and it is given by:
\begin{equation*}
	A=\frac{I_A-I_B}{I_A+I_B}
\end{equation*}
where $I_{A}$ and $I_{B}$ are the intensities measured using opposite magnetizations of the FeO film. 
The spin polarization $P$ is related to the spin asymmetry of the scattered electrons:
\begin{equation*}
	P=\frac{A}{S}=\frac{I_{\uparrow}-I_{\downarrow}}{I_{\uparrow}+I_{\downarrow}}
\end{equation*}
where $I_{\uparrow}$ and $I_{\downarrow}$ are the spin populations of opposing spins where $S$ 
denotes the Sherman function that was equal to 0.18 for the data shown 
in Fig.~2 of the main text, and 0.25 for the data shown 
in Fig.~3 of the main text and Fig.~\ref{spin_z}. 
Finally, the spin populations $I_{\uparrow}$ and $I_{\downarrow}$ can be written as
\footnote{These are not the absolute, but rather relative intensities with arbitrary units 
since the electron reflectivity of the $Fe$ plate is not taken into account.}:
\begin{equation*}
	I_{\uparrow}=(1+P)\frac{I_A+I_B}{2}   \>\>\>\>  \textmd{and}  \>\>\>\>  I_{\downarrow}=(1-P)\frac{I_A+I_B}{2}
\end{equation*}
The energy (or momentum) dependence of $I_{\uparrow}$ and $I_{\downarrow}$ 
yield the spin-resolved energy (or momentum) distribution curves. 
The typical energy resolution of the spin-resolved EDCs was 30~meV. 
The angular step of the spin-resolved MDCs was 0.5$^{\textmd{o}}$. 
The polarization of incoming photons was linear horizontal 
and all measurements were performed at 25~K. 

\subsection*{Energy-momentum dispersion around $\bar{\textmd{M}}$}
\begin{figure*}[t!] 
   \centering
   \includegraphics[clip, width=0.95\textwidth]{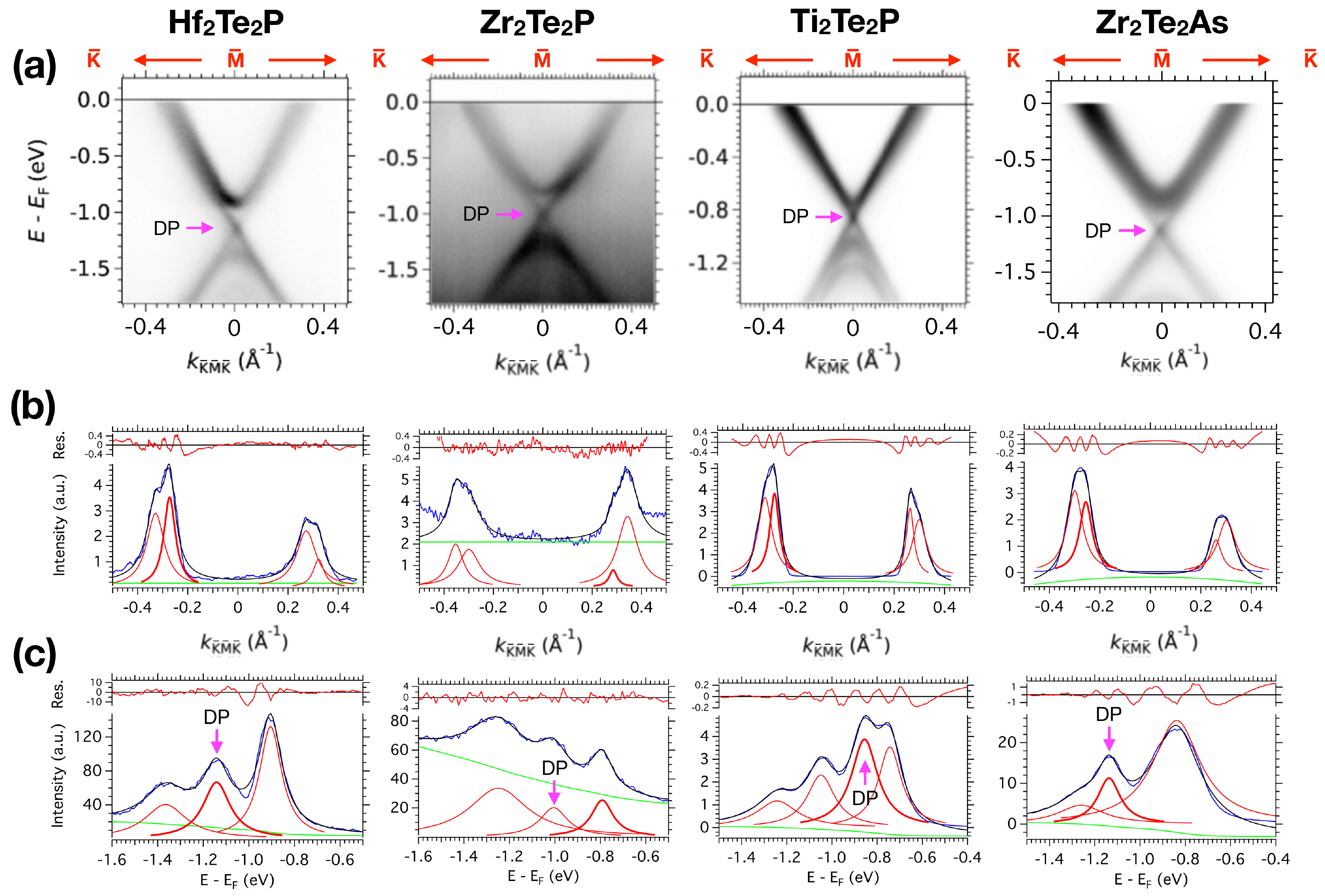} 
   \caption{(a) Energy-momentum dispersion along $\overline{\textmd{KMK}}$ 
   			in Hf$_2$Te$_2$P, Zr$_2$Te$_2$P, Ti$_2$Te$_2$P and Zr$_2$Te$_2$As. 
   			The purple arrows point to the Dirac points (DPs) of linearly dispersing 
   			topological surface states. 
   			(b) Momentum distribution curves (MDCs) at $E_F$ (integrated over $\pm$10~meV) 
   			of states around $\bar{\textmd{M}}$ in panel (a). 
   			(c) Energy distribution curves (EDCs) at $\bar{\textmd{M}}$ (integrated over $\pm$0.005~\AA$^{-1}$). 
   			These MDCs and EDCs (blue curves) are fitted with Lorentzian peaks (red peaks) 
   			and a quadratic (MDC) or Shirley (EDC) background (the green curve). 
   			The overall fits are the black curves, while the red line in each 
   			top inset is the fitting residual.}
   \label{EK_M}
\end{figure*}

Fig.~\ref{EK_M} (a) presents the linearly dispersing Dirac states 
along the $\overline{\textmd{KMK}}$ direction -in the vicinity of $\bar{\textmd{M}}$- 
for Hf$_2$Te$_2$P, Zr$_2$Te$_2$P, Ti$_2$Te$_2$P and Zr$_2$Te$_2$As. 
From the MDCs integrated around the Fermi level and the following Lorentzian peak fits 
shown in Fig.~\ref{EK_M} (b), we can distinguish the Fermi momenta 
of these topological surface states from those of the bulk conduction bands. 
This is clear experimental proof that the topological states at $\bar{\textmd{M}}$ are metallic. 
From the EDCs integrated around $\bar{\textmd{M}}$ and the corresponding Lorentzian peak fits 
shown in Fig.~\ref{EK_M} (c), the binding energies $E_D$ of the Dirac points (DPs)
and the bulk band-gaps $E_{gap}$ can be extracted. 
The obtained values are listed in Tab.~\ref{MDC_EDC_num}.

\begin{table}[h]
\centering
\begin{tabular}{c|c|c|c|c}
\hline
\hline
 & Hf$_2$Te$_2$P & Zr$_2$Te$_2$P & Ti$_2$Te$_2$P  & Zr$_2$Te$_2$As \\
\hline
$E_D \pm 0.01$ (eV)& 1.14 & 1.01 & 0.85 & 1.14  \\
\hline
$E_{gap} \pm 0.01$ (eV)  & 0.48 & 0.45 & 0.31 & 0.43  \\
\hline
\hline
\end{tabular}
\caption{Typical quantities characterizing the electronic structures of M$_2$Te$_2$X 
		(with M = Ti, Zr or Hf, and X = P or As), extracted from the MDCs and EDCs 
		quantitative analysis. $E_D$s are the binding energies of the Dirac points 
		at $\bar{\textmd{M}}$. $E_{gap}$s are the bulk bandgaps at $\bar{\textmd{M}}$.}
\label{MDC_EDC_num}
\end{table}

At $\bar{\textmd{M}}$, the spin-orbit interaction opens a gap at the crossing point 
of the parent bulk bands that are mainly composed of Te $p$-bands, 
and drives a band inversion inducing a topological surface state in the gap opening. 
Along the $\overline{\Gamma \textmd{\textmd{M}} \Gamma}$ direction, 
these bulk Te $p$-bands connect with the bulk bands in the vicinity of $\bar{\Gamma}$ 
(Fig.~\ref{ZTP_HTP_ZTA_arc}). The image sequence in the bottom panels 
of Fig.~\ref{ZTP_HTP_ZTA_arc} can also track the variations in the size 
of the energy gap and the energy position of the Dirac point as it will be explained 
in the next subsection. The values summarized in Tab.~\ref{MDC_EDC_num} 
can be related with variations in the strength of the atomic spin-orbit coupling (SOC) 
among the compounds of the tetradymite family M$_{2}$Te$_{2}$X 
(with M$=$Ti, Zr or Hf and X$=$P or As). 
SOC is systematically increased by replacing Ti with Zr and finally with Hf,
as well as P with As.
We may therefore conclude that such an increase 
of the atomic SOC by elemental substitution is reflected into an increase in $E_{gap}$
(except for the replacement P$\rightarrow$As, which keeps the gap essentially unaltered),
and $E_D$.
The increase in $E_D$ might be also related to a slight doping caused by
defects.

\subsection*{Dirac-Node arcs}
\begin{figure*}[t!] 
   \centering
   \includegraphics[clip, width=0.95\textwidth]{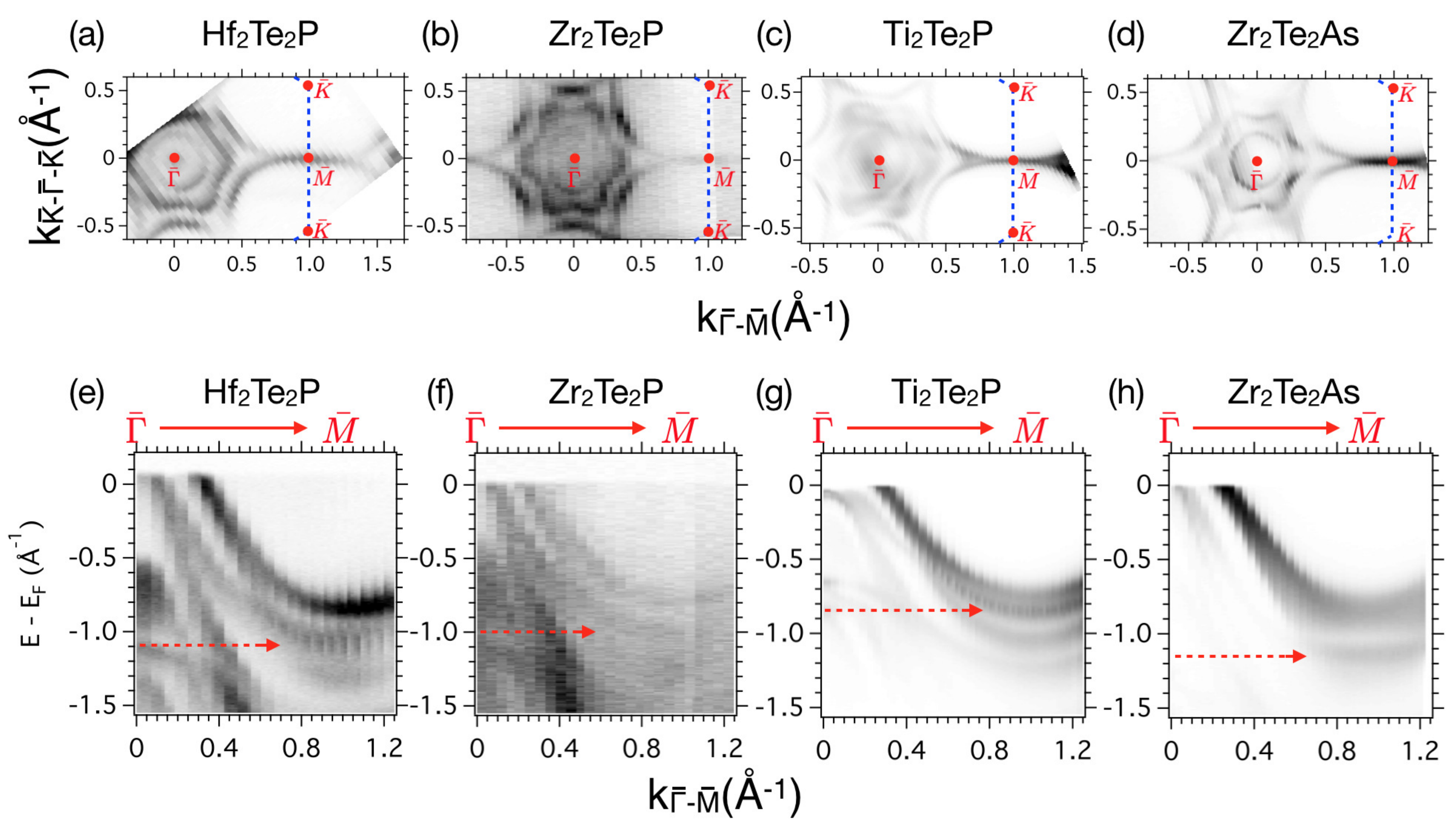} 
   \caption{(a-d) Constant energy maps around the Dirac point 
   			of Hf$_2$Te$_2$P, Zr$_2$Te$_2$P, Ti$_2$Te$_2$P and Zr$_2$Te$_2$As 
   			(hence at binding energies 1.14~eV, 1.00~eV, 0.85~eV and 1.15~eV, respectively), 
   			together with the reduced 2D first Brillouin zones (blue dashed lines). 
   			(e-h) The electronic dispersions along $\overline{\Gamma \textmd{\textmd{M}}}$. 
   			The red arrows indicate the quasi-flat states leading 
   			to the corresponding Dirac-node arcs. 
   			All the data were acquired with LH polarized photons 
   			at temperatures $T~=~25$~K (a,~b,~e,~f) 
   			and $T~=~6$~K (c,~d,~g,~h).}
   \label{ZTP_HTP_ZTA_arc}
\end{figure*}

Fig.~\ref{ZTP_HTP_ZTA_arc} (a-d) present the constant energy maps 
of Hf$_2$Te$_2$P, Zr$_2$Te$_2$P, Ti$_2$Te$_2$P and Zr$_2$Te$_2$As, 
at binding energies 1.14~eV, 1.00~eV, 0.85~eV and 1.15~eV respectively. 
These energy values correspond to the energy position of the Dirac point 
in each different compound. The energy contours show the presence of Dirac-node arcs. 
Fig.~\ref{ZTP_HTP_ZTA_arc} (e-h) show the corresponding energy-momentum dispersion 
along the Dirac-node arcs, i.e. $\overline{\Gamma \textmd{\textmd{M}}}$ direction. 
The dashed red arrows indicate the binding energies of the Dirac crossing points at $\bar{\textmd{M}}$. 
The pointed band bottoms, corresponding to the Dirac-node arcs, are relatively flat, i.e. massive. 
We note again that, from Hf, to Zr, and to Ti, the strength of spin-orbit coupling (SOC) 
is systematically reduced. As a direct result, the binding energies of the Dirac point decrease. 
The binding energy of the Dirac point at $\bar{\textmd{M}}$ in Zr$_2$Te$_2$As, 
whose SOC strength is larger than Zr$_2$Te$_2$P, 
behaves consistently with the aforementioned observation. 

\subsection*{Out-of-plane spin polarisation and total amplitude of the spin polarisation}

\begin{figure*}[t!]
   \centering
   \includegraphics[clip, width=0.95\textwidth]{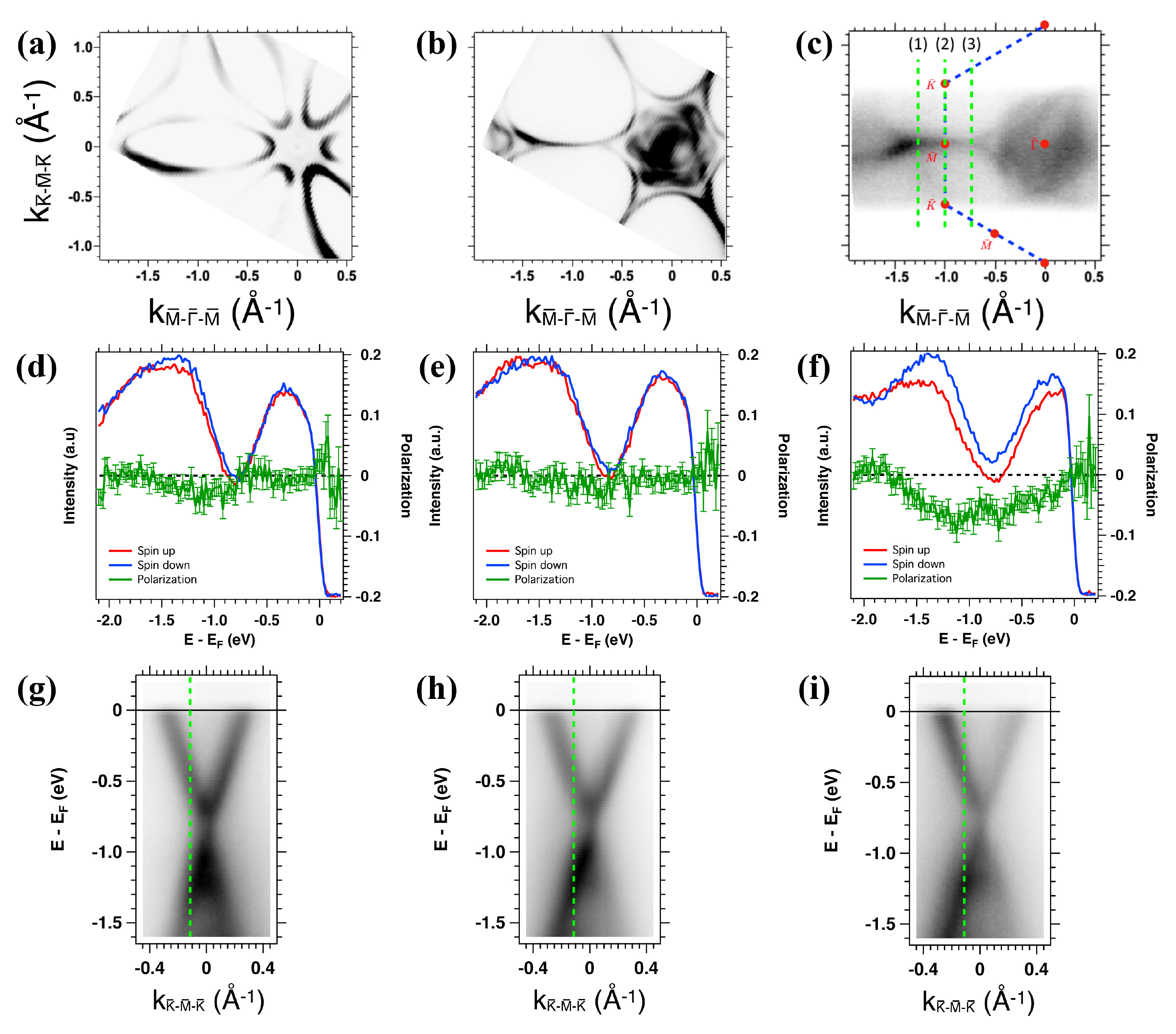} 
   \caption{\textbf{Out-of-plane spin polarization along the Dirac node arc 
   			of Ti$_{2}$Te$_{2}$P.} 
   			(a), (b) Spin-integrated ARPES results obtained at synchrotron SOLEIL 
   			with LH polarized photons of 50~eV at a temperature of 5~K 
   			showing the Fermi surface contours of Ti$_{2}$Te$_{2}$P (a) 
   			and the existence of linear features at a binding energy of 0.85~eV (b). 
   			(c) Same as panel (b) but measured with LH photons of 55~eV 
   			at the spin-resolved ARPES setup of beamline 9B at HiSOR. 
   			Path (1) corresponds to data shown in panels (d) and (g), 
   			path (2) to panels (e) and (f), and path (3) to panels (f) and (i). 
   			(d)-(f) Spin-resolved EDCs (red, blue) and energy dependent 
   			out-of-plane spin polarization along $\overline{\Gamma \textmd{\textmd{M}}}$ (green) 
   			measured at the left branch of the Dirac cones [panels (g)-(i)], 
   			themselves obtained at different $k$-locations along the Dirac node arc. 
   			(g)-(i) Energy-momentum dispersion along the $k$-paths indicated in panel (c), 
   			with overlaid vertical dashed lines showing the $k$-location 
   			of the spin-resolved EDCs shown in panels (d)-(f). 
   			Panels (c)-(f) were all measured with LH photons of 55~eV
   			at a temperature of 25~K.}
   \label{spin_z}
\end{figure*}

Fig.~\ref{spin_z} presents the out-of-plane component of the spin polarization 
along the Dirac-node arc of Ti$_{2}$Te$_{2}$P, measured in the same way as Fig.~3 
in the main text. As required by symmetry, at $\bar{\textmd{M}}$, 
the out-of-plane component of the spin polarization is, within experimental accuracy, 
zero -see Fig.~\ref{spin_z}(e). 
However, as seen in Figs.~\ref{spin_z}(d,~f), a small (non-zero) 
negative out-of-plane spin polarization can be observed away from the $\bar{\textmd{M}}$ point,
but still on the Dirac-node arc, for the states comprising its upper and lower cones. 

The observed value of the in-plane spin polarization, as shown in the main text, is smaller than 1. 
The magnitude of the total 3D spin polarization vector would be also below unity, 
even if we take into account its even smaller out-of-plane projection (Fig.~\ref{spin_z})
-which, as required by symmetry, vanishes at the $\overline{\textmd{\textmd{M}}}$ point 
within experimental accuracy.
The experimental observation of 
spin-polarized electrons that are less than 100\% polarized is, in fact, not surprising. 
Similar values of incomplete spin polarization have been previously measured on Bi-based 
topological insulators \cite{Hsieh2009}, thin films 
and single crystals of Sb and Bi \cite{Hsieh2010, Takayama2014, Hirahara2007} 
and Rashba systems with giant splitting \cite{Meier2008, Yaji2010}. 
According to those works, reduced values of spin polarization can be due to the overlap 
of adjacent photoemission peaks with opposite polarization \cite{Meier2008, Hsieh2009, Hsieh2010} 
and/or because of the immediate vicinity of the surface states 
with non-polarized bulk bands \cite{Yaji2010, Hirahara2007, Takayama2014}. 
Indeed, for all M$_{2}$Te$_{2}$X compounds 
the upper branch of the Dirac cone at $\overline{\textmd{\textmd{M}}}$ lies very close to the projected bulk bands 
that form the petal contours at the Fermi surface, while its lower branch lies close to 
multiple hole-like states of bulk or surface origin \cite{Chen2018}. 
Moreover, we point out the fact that spin polarization of the photoelectrons in an ARPES experiment 
can be strongly influenced by the, experimental geometry, photon polarization 
and photon energy~\cite{Jozwiak2011, Jozwiak2013, Barriga2014}. 
In addition, the limited energy and angular resolution of the experiment 
would cause the reduction of observed spin-polarization.
As a consequence, the exact magnitude of the inherent spin polarization 
cannot be determined accurately.

\subsection*{Tuning the Dirac points by potassium deposition on the sample's surface}

\begin{figure}[t!h]
   \centering
   \includegraphics[clip, width=0.95\linewidth]{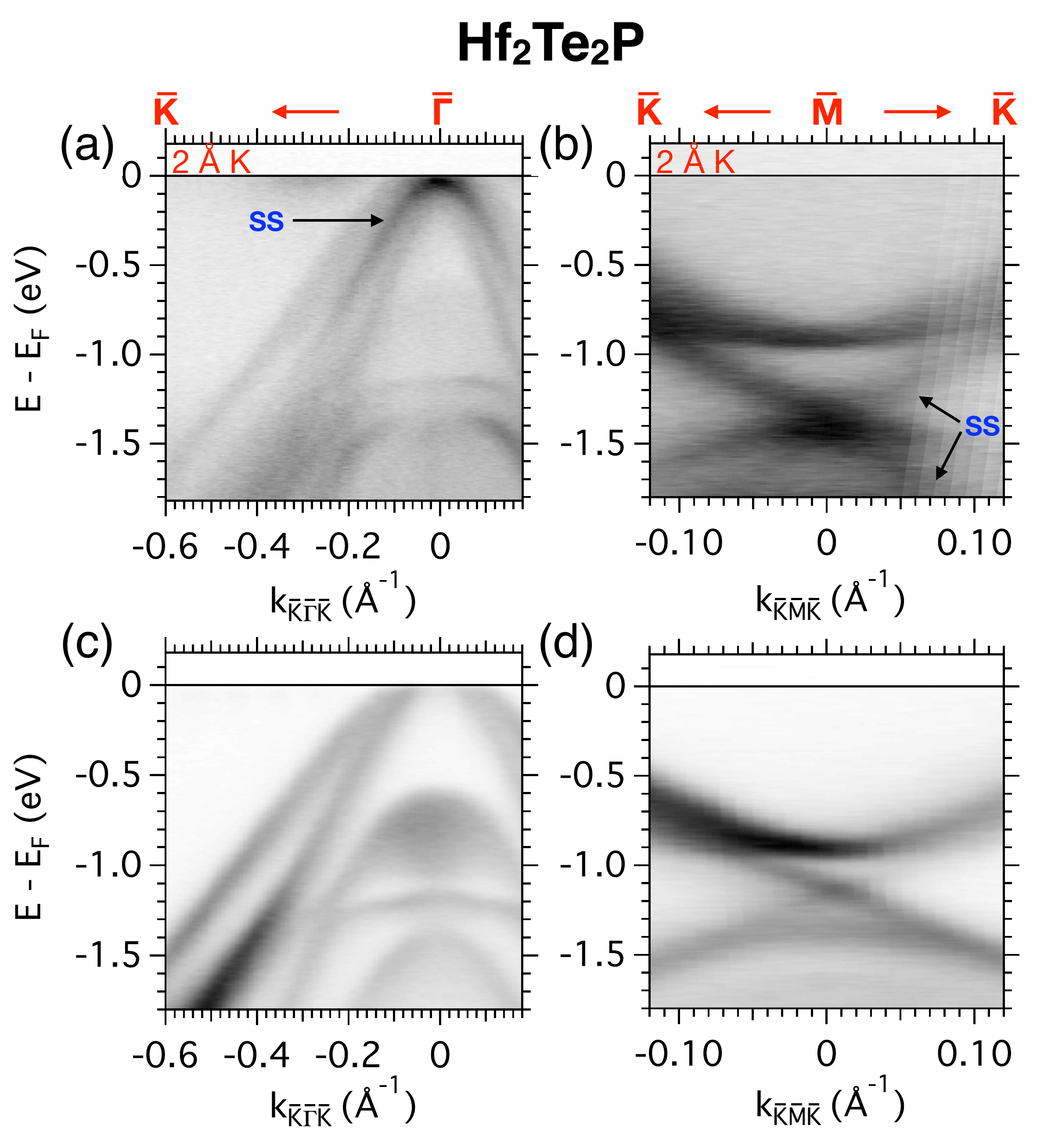} 
   \caption{(a,~b) ARPES energy-momentum intensity maps at the surface of Hf$_2$Te$_2$P 
   			capped with 2~\AA~of potassium, measured respectively at 
   			$\bar{\Gamma}$ along $\bar{\textmd{K}}-\bar{\Gamma}-\bar{\textmd{K}}$ 
   			and at $\bar{\textmd{M}}$ along $\bar{\textmd{K}}-\bar{\textmd{M}}-\bar{\textmd{K}}$. 
   			(c,~d) Analogous measurements at the bare (i.e., non capped) 
   			surface of Hf$_2$Te$_2$P. 
   			All the data were acquired with LH polarised photons 
   			at an energy $h\nu=50$~eV at $T=20$~K. 
   			The surface state (SS) is indicated by the black arrows in panel (b).
   			}
   \label{HTP_K_states}
\end{figure}

For real applications, it would be desirable to have the Dirac point at $E_F$ 
such that the backscattering in the solid will be strongly suppressed, 
leading to dissipation-less transport. Note that, according to the DFT calculations, 
Hf$_2$Te$_2$P has an additional Dirac surface state around $200$~meV above the Fermi level 
at $\bar{\Gamma}$~\cite{Chen2018, Hosen2018}, which could even be around $100$~meV 
lower in energy given the observed energy shift between the ARPES data and the calculations. 
These considerations motivated us to try to tune the Fermi level by potassium (K) 
deposition at the sample's surface, thus hoping to bring the Dirac point of Hf$_2$Te$_2$P 
within the energy range of the occupied bands. 
The potassium deposition could also serve to experimentally distinguish 
the surface states from the bulk states.

Fig.~\ref{HTP_K_states}(a) shows the electronic states at $\bar{\Gamma}$ 
along $\bar{\textmd{K}} - \bar{\Gamma} - \bar{\textmd{K}}$ measured at the surface of 
Hf$_2$Te$_2$P capped with 2~\AA~of potassium. 
Compared to the bare surface of Hf$_2$Te$_2$P in Fig.~\ref{HTP_K_states}(c), 
a new hole-like surface state with its top energy right below the Fermi level, 
indicated with an arrow, appears around $\bar{\Gamma}$. 
On the other hand,  there are no observable energy shifts of the bulk bands 
except that the signal-to-noise ratio is much lower due to K capping. 

Fig.~\ref{HTP_K_states}(b) shows the electronic states including the Dirac surface state 
at $\bar{\textmd{M}}$ along $\bar{\textmd{K}} - \bar{\textmd{M}} - \bar{\textmd{K}}$ 
measured after the K capping at Hf$_2$Te$_2$P surface. 
Compared to Fig.~\ref{HTP_K_states}(d), the Dirac crossing point of the surface state (SS) 
is shifted down by around 200~meV to a binding energy 1.35~eV, 
while the bulk bands (above and below the gap) remain unchanged. 

The fact that surface electron doping (K deposition) shifts only the Dirac cones
is an important observation that proves the surface nature of such topological states.

\subsection*{Orbital contribution and band connectivity analysis based on slab DFT calculations}
\begin{figure}[t!h] 
   \centering
   \includegraphics[clip, width=0.95\linewidth]{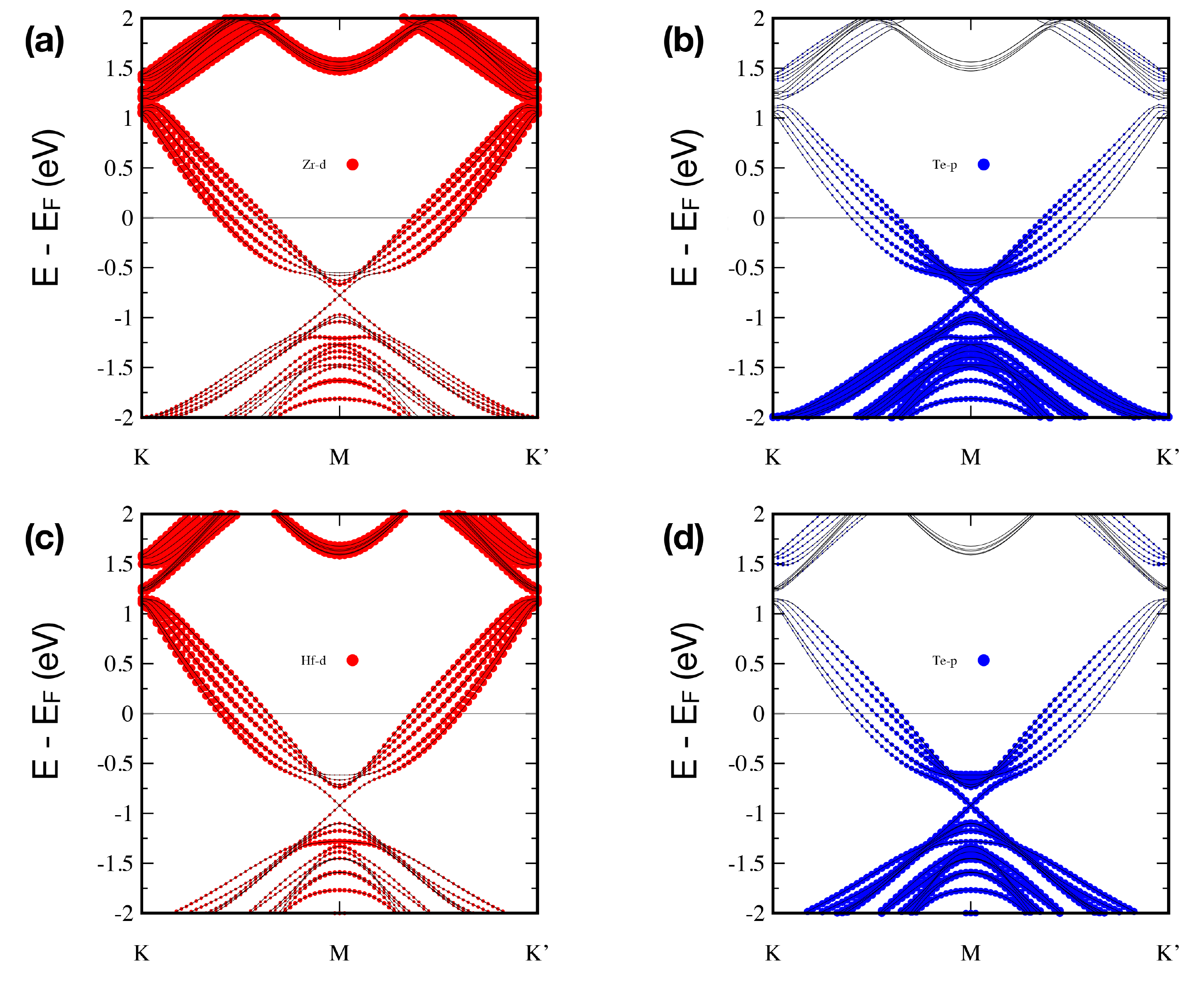} 
   \caption{Slab band dispersion along $\bar{\textmd{K}} - \bar{\textmd{M}} - \bar{\textmd{K}}$  
   			high symmetry direction along with the orbital character of the bands: 
   			(a, b) Zr-$d$ and Te-$p$ orbital resolved bands for Zr$_2$Te$_2$P; 
   			(c, d) Hf-$d$ and Te-$p$ orbital resolved bands for Hf$_2$Te$_2$P. 
   			The size of the dots is proportional to the percentage of the orbitals 
   			mentioned in the figure.
   			}
   \label{orbit_types}
\end{figure}

\begin{figure}[t!h] 
   \centering
   \includegraphics[clip, width=0.95\linewidth]{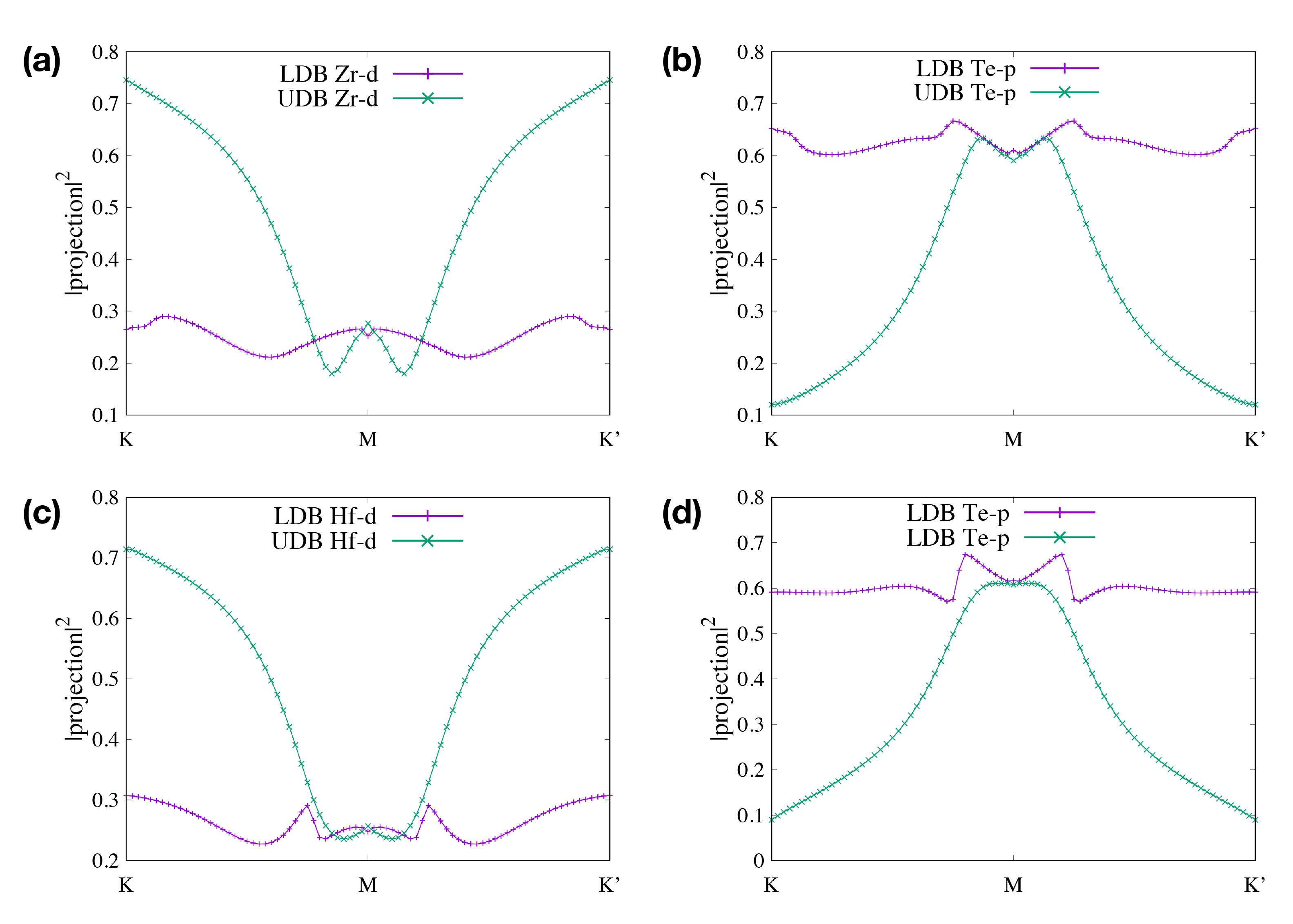} 
   \caption{Quantitative comparison of the orbital composition of the bands 
   			contributing to the surface Dirac cone state at $\overline{\textmd{M}}$ point: 
   			(a, b) Zr$_2$Te$_2$P;
   			(c, d) Hf$_2$Te$_2$P. 
   			LDB: lower Dirac branch; UDB: upper Dirac branch.
   			}
   \label{orbit_portions}
\end{figure}

\begin{figure}[t!h] 
   \centering
   \includegraphics[clip, width=0.95\linewidth]{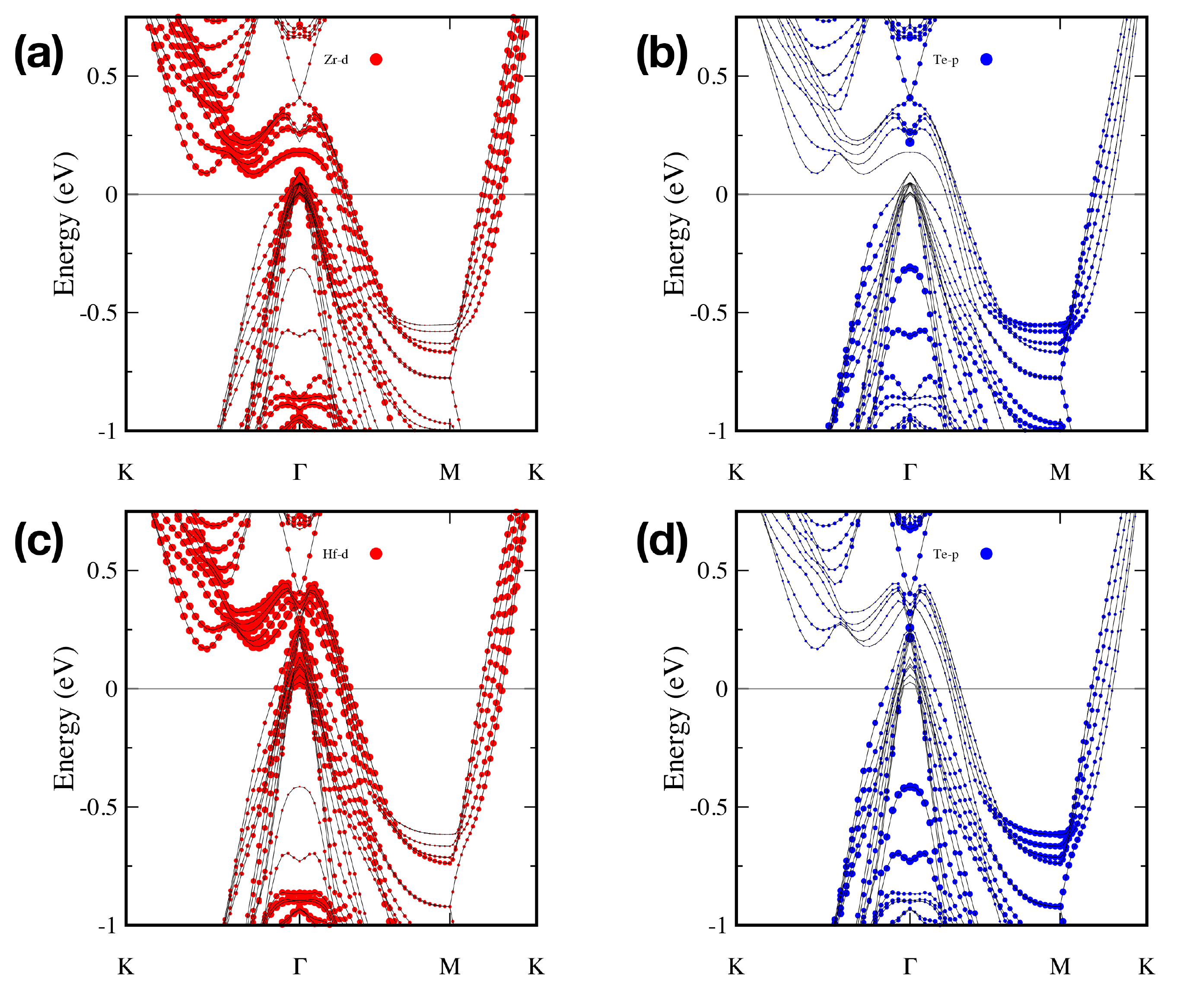} 
   \caption{Orbitally resolved slab band dispersion 
   			along $\overline{\textmd{K}}-\overline{\Gamma}-\overline{\textmd{M}}-\overline{\textmd{K}}$
   			direction for (a, b) Zr$_2$Te$_2$P and (c, d) Hf$_2$Te$_2$P.
   			}
   \label{band_connectivity}
\end{figure}

We performed slab DFT calculations of all members of the M$_{2}$Te$_{2}$X family 
and found that there is no significant differences in the orbital character 
of the surface Dirac cones formed at the $\overline{\textmd{M}}$ point 
between the strong  and the weak topological metal members of the family.

In Fig.~\ref{orbit_types} and~\ref{orbit_portions}, we present the results 
of slab DFT calculations in order to compare the orbital contributions 
to the Dirac states in the tetradymite compounds with strong (i.e. Zr$_2$Te$_2$P and Ti$_2$Te$_2$P) 
and weak (i.e. Hf$_2$Te$_2$P) topological character. 
In all panels, it is evident that both $p$ (Te) and $d$ (Zr, Hf or Ti) orbitals 
contribute to the gapless surface state at the $\overline{\textmd{M}}$ point, 
which may explain the strong anisotropy of the surface states at $\overline{\textmd{M}}$, 
in line with the case of Ru$_2$Sn$_3$ \cite{gibson2014quasi}. 
The corresponding percentage right at the $\overline{\textmd{M}}$ point 
is of the order of 60 to 70\% for both Zr$_2$Te$_2$P and Hf$_2$Te$_2$P 
and stays relatively constant for the upper part of the Dirac cone 
as one moves towards $\overline{\textmd{K}}$. 
On the other hand, the lower part of the Dirac cone changes progressively 
to a predominant Zr (or Hf) $d$ character as one approaches $\overline{\textmd{K}}$. 
Most importantly, only minor differences in the orbital composition are observed 
between weak and strong topological metals showing that orbital composition 
cannot be the key to the different topological character of these compounds.

Our slab calculations in Fig.~\ref{band_connectivity} reveal the detailed dispersion 
and the connectivity of the electron bands for both a strong (i.e. Zr$_2$Te$_2$P) 
and a weak (i.e. Hf$_2$Te$_2$P) topological metal member of the M$_2$Te$_2$X family. 
These calculations are in excellent agreement with the massive (almost flat) 
dispersion of the Dirac-node arc from $\overline{\textmd{M}}$ to $\overline{\Gamma}$. 
Furthermore, they -once more- confirm Zr$_2$Te$_2$P (Hf$_2$Te$_2$P) 
as a strong (weak) topological metal by predicting the absence (existence) 
of a gapless state at $\overline{\Gamma}$ right above the Fermi level. 
Notice that for the Zr-compound, Fig.~\ref{band_connectivity}(a), 
there is only one Dirac cone at around $0.4$~eV, while for the Hf-compound, 
Fig.~\ref{band_connectivity}(c), there are two Dirac cones, one at around $0.4$~eV 
and the other at around $0.25$~eV. 
Notice also that for Hf$_2$Te$_2$P the Dirac-node arc does indeed connect 
with the gapless state at $\Gamma$. 
However, the band connectivity is rather complicated as the surface bands 
of predominant d character in the Hf-compound cross an even number of times, 
while in the Zr-compound they cross an odd number of times. 
Most importantly, the absence of a gapless state at $\Gamma$ 
does not affect the presence of the Dirac-node arc.

%
\end{document}